# Exponentially Damped Breit-Pauli Hamiltonian for the Description of Positronium Decay and Other High-energy Processes

Robert J. Buenker*

Bergische Universität – FBC-Mathematik und Naturwissenschaften

Gaußstr. 20, D-42119 Wuppertal , Germany

**Abstract**

The conventional explanation for the 1.022 MeV decay of positronium in terms of the annihilation of an electron and its antiparticle is questioned because of the impossibility of proving experimentally that matter actually disappears. Recent work has provided evidence that photons are not destroyed in radiative absorption, for example, but rather attain an undetectable (E=$h\nu$=0) state. Questioning the creation-annihilation hypothesis in the case of positronium decay leads to the construction of an exponentially damped Breit-Pauli Hamiltonian which when employed in a Schrödinger equation gives an $e^+e^-$ ground state with a binding energy of 1.022 MeV. The analogous treatment does not produce any (unobserved) states of similarly low energy for the hydrogen atom, but by virtue of a simple scaling theorem it does lead to a maximal proton-antiproton binding energy of 1.876 GeV, which is exactly $m_p/m_e$ times larger than for $e^+e^-$. Arguments are presented to identify the resulting tightly bound $e^+$ $e^-$ system with the photon itself, since both are ascribed a zero rest mass. The above Hamiltonian is identified with the unified electroweak interaction, especially since it also proves applicable to the description of neutrino processes. It is pointed out that the analogous $\bar{\nu}\nu$ binary systems would be subject to predissociation into free neutrinos and antineutrinos, however, unlike their much more strongly bound $e^+$ $e^-$ and $p^+$ $p^-$ counterparts. This observation suggests a different explanation for the well-known solar neutrino problem than the flavor-oscillation theory which has been proposed earlier.

# I. Introduction

After the discovery of quantum mechanics, nothing quite so excited the imagination of physicists as Dirac's prediction [1] of the existence of antimatter. Within a few years Anderson had demonstrated [2] the existence of the positron as the antiparticle of the electron. and within decades the antiproton [3] had also been found. Although Dirac's line of reasoning involved first and foremost his treatment [1] of the fine structure in the spectra of hydrogenic atoms, the ultimate theoretical basis for his prediction can be traced back years before to Einstein and the special theory of relativity [4]. The famous mass-energy equivalence relation, $E=mc^2$, was given a more concrete interpretation, namely that two particles of equal rest mass but opposite electric charge could be converted entirely into energy. In the case of $e^+$ and $e^-$ the energy appears as electromagnetic radiation or photons, whose frequency $\nu$ is related to the released energy by the Bohr relation [5], $E=h\nu$.

The creation and annihilation of matter, as this phenomenon is generally called, is central to the theory of modern physics, ranging from the submicroscopic regime of atoms, electrons, photons and more exotic elementary particles to the supermacroscopic world comprising astronomy and cosmology. Indeed one uses these terms to discuss processes that are commonplace in our everyday experience, such as the absorption and emission of light. The energy released or gained upon transitions between different states in atoms or molecules is said to cause the creation or destruction of a photon, again in accordance with the Bohr frequency relation.

Recently [6], attention has been called to the fact that by virtue of the relativistic Doppler effect photons of arbitrarily small kinetic energy can have their frequency modified to experimentally detectable values by having the observer approach the source of the radiation at extremely high speeds. This analysis proves that annihilation does not occur for photons of such low energy. But if photons of infinitesimally small kinetic energy exist, it seems inconsistent to claim that those at the limit of exactly zero energy do not exist as well. Moreover, it is possible to associate the vacuum field of the theory of quantum electrodynamics with the existence of a uniformly high density of (real) zero-energy photons throughout the universe [6] rather than with the occurrence of creation-annhilation processes. In considering these aspects of photon theory, one should be aware of the fact that the Bohr frequency relation [5] indicates that a group of photons of exactly zero energy must have a null frequency and an infinitely long wavelength, thereby precluding their observation by spectroscopic means. This situation is independent of the state of motion of the observer as well, again as can be inferred from the relativistic Doppler effect.

There is thus a clear alternative to the creation and annihilation hypothesis in the interpretation of radiative absorption and emission processes. We live in a sea of photons, most of which have exactly zero kinetic energy. In reality these photons simply pass between an undetectable and a detectable state as they either gain or lose energy because of a transition of an atom or molecule. It only *appears* that they are either created or annihilated in the process. With this background in mind, it is interesting to reconsider the interpretation of other physical processes in which it is assumed that matter is either created or destroyed. Is it possible, for example, that when an electron and positron interact, they also assume an undetectable state without actually going out of existence [2]? Especially since photons are observed as the product of this interaction, there is a clear basis upon which to begin such a discussion.

**II. Positronium Decay Without Creation and Annihilation of Matter**

When an electron and positron combine the initial product is positronium but, after a short time, high-energy photons are produced and all trace of the original two particles is lost. Since the amount of energy carried away by the photons is exactly equal to the rest-mass equivalent of positronium (1.022 MeV), it is natural to interpret this process as involving the complete destruction of the original two particles. In general, the more exothermic such a process, the more intuitively reasonable such an explanation becomes. Yet it cannot be proven with certainty that the electron and positron pass out of existence, because it is impossible to eliminate the possibility that a new state has been reached which cannot be detected by experimental means. Under the circumstance, one is justified to make the opposite assumption and see if it leads to any sort of contradictory result.

In the present case, this procedure forces one to assume that even in positronium decay there is elemental balance, in particular that the electron and positron at the start of the reaction are still present in a real form at its conclusion as well. Consistency with the arguments given in the Introduction requires one to assume that the photons produced at the end of the decay process were present in a massless and therefore undetectable form at its beginning, so there is no new difficulty provided by this aspect of the experimental observations. The situation with the electron and positron is less clear, however, because they have a combined rest mass equivalent to 1.022 MeV and are already in their lowest translational state at the beginning of the process.

To proceed further it is helpful to recall that there is a close analogy between the weakly bound positronium $e^+ e^-$ system and the hydrogen atom. In quantum electrodynamics (QED) it is easily shown that the two systems possess a very similar set of internal states whose respective binding energies differ by a factor of very nearly one-half, the ratio of the positronium reduced mass to that of the hydrogen atom. Nevertheless, the experiment under discussion emphasizes that there is a crucial distinction between these two systems when it comes to occupying their 1s states. For the hydrogen atom this state is stable indefinitely, whereas for positronium it is decidedly unstable, with a lifetime of only $10^{-10}$ s in the case of the J=0 multiplet. Both systems can undergo spontaneous radiative decay from their respective 2p states, and this process is well known to involve a dipole-allowed transition to the lower 1s state of each system. The fact that no subsequent emission has ever been observed from the hydrogen $^2S_{1/2}$ state is why one refers to it as the ground state of this system.

But does this mean that the analogous 1s state of positronium is its ground state? The close similarity between the hydrogen atom and positronium spectra predicted by quantum mechanical theories ranging from the non-relativistic Schrödinger equation to QED is the primary justification for answering this question in the affirmative, but the observed positronium decay from its 1s state raises the possibility that this conclusion is incorrect. Indeed, since we have begun this discussion by assuming that the electron and positron of this system remain in existence at the conclusion of the process, there is really no choice in the present context but to assume that there must be another state of the $e^+$ $e^-$ system lying far below its 1s counterpart, and that the photons emitted in the decay result because of a transition between these two states. At the same time, the fact that no comparable emission process from its 1s state occurs for the hydrogen atom clearly suggests that there is no such lower-lying state for this system.

The appearance of two or more photons in the case of positronium decay but only one in conventional atomic emission processes does not conflict in any way with this interpretation. Such a distinction is easily understandable in terms of energy-momentum conservation arguments. In low-energy transitions such as the 2p→1s example, part of the energy released is carried away by the atom itself. Its translational momentum must simply be equal and opposite to that of the emitted particle(s) and this condition is easily met by a single photon. The fact that the amount of energy given off in positronium decay is exactly equivalent to the rest mass of the initial (weakly bound) system demands in the present interpretation that the mass of the final $e^+e^-$ state be exactly zero. As a result, regardless of whether one assumes that the original electron and positron have been annihilated in the positronium decay process or have instead entered into a new state of much higher binding energy, it is clear that the corresponding product system cannot carry away momentum to oppose that of any emitted photons. Under the circumstances, the only way to satisfy the conservation laws in this case is to have the energy produced in the decay to be carried away by two or more photons, as is observed.

The massless nature of the $e^+$ $e^-$ final state in the above interpretation suggests something else. There is already another system with E=0 involved in this general discussion, namely the photon. The possibility thus arises that they are in fact one and the same. In other words, the tightly bound $e^+e^-$ state lying 1.022 MeV lower in energy than the positronium 1s state could be identical with the photon itself (Fig.1). Such an identification is also needed if the vacuum field of QED theory is to be associated with real zero-energy photons [6], since positron-electron pairs must be assumed to be formed from them in order to maintain the elemental balance required in the present interpretation.

With the help of this assumption, the creation and annihilation of matter hypothesis can again be avoided and the following picture of the positronium decay process results (Fig.2). In addition to the three systems actually observed in the course of the entire process (for J=0 decay), namely positronium in its 1s state at the outset and two photons of 0.511 MeV energy each at its completion, there are also three undetected photons with E=0. The new species are identical to one another, with two of them present at the start of the decay process (pre-emission photons) and the other being the final $e^+e^-$ state itself (akin to an absorption photon after it has lost all its energy).

In this way, positronium decay is seen to be closely related to an ordinary emission process in which an atom or molecule undergoes a transition to a state of lower energy (Fig.1). In this case, however, the amount of energy lost is exactly equal to the rest mass of positronium multiplied by the square of the velocity of light, and hence the final state has no mass. Thus the process more nearly resembles one of nuclear binding, except that in this case all of the system's original mass is lost rather than just a small portion of it. The other initial reactants are two E=0 photons present in the immediate neighborhood (Fig.2), as discussed in the previous section. They share the energy lost in the transition in exactly equal proportions by virtue of the requirement that energy and linear momentum both be conserved in any such process, again as discussed above.

To further investigate this interpretation of the positronium decay process, it is important to consider two related questions. First of all, does the hypothesis of an $e^+e^-$ structure for the photon lead to any contradiction in terms of experimental observations? Secondly, what kind of interaction is capable of binding an electron and its antiparticle so strongly together that it produces the total loss of mass observed? This aspect is particularly interesting since one knows that no comparable state exists for the hydrogen atom. On the other hand, the proton and antiproton combine to produce 1836 times as much energy, so one would like to find an interaction which would explain this process in a very similar manner as for positronium decay. In general, it can be seen as an advantage for the present approach that it even raises the question of what forces are ultimately responsible for the large exothermicities of particle-antiparticle interactions. Such an explanation is notably absent in the creation-annihilation interpretation, in which it is simply noted that the energy lost is exactly equivalent to the mass of the reactant species. There is a measureable decrease of mass in nuclear reactions as well, but considerable effort has gone into characterizing the forces which bind the nucleons together and thereby produce this result. A parallel development would be welcome in the present case, but it is understandably difficult to approach this question when it is assumed that the original particles simply disappear instead of becoming quite strongly bound to one another.

Before considering these two questions in the following sections, however, it is well to clear up another issue regarding the present interpretation, namely why does the electron-positron system always end up in a state of exactly zero energy as a result of the proposed transition (Figs. 1-2), The same question arises in conjunction with the interpretation of radiative absorption processes mentioned in the Introduction. In a companion paper [6], it is argued that the special properties of the E=0 photon states indicated by the Einstein mass dilation formula, namely the capacity to move at any speed less than c, down to and including v = 0, hold the key to this puzzle. As for the interaction of radiation with atoms and molecules, it seems essential that the $e^+$-$e^-$ reaction occur in a narrow region of space-time to be at all feasible. If the product photon (or $e^+e^-$ species in the present interpretation) has any kinetic energy at all, it must fly away from the reactive site with the speed of light, making it impossible to fulfill this condition in a satisfactory manner. Thus the proposed null rest mass of the final $e^+e^-$ state is seen to be quite consistent with the observed findings, even though at first glance it might appear otherwise.

In summary, in lieu of finding satisfactory answers to the two questions posed above, it should be clear that a way has been found that is consistent with the arguments employed earlier to describe radiative absorption and emission processes which is capable of explaining the positronium decay process without relying on the hypothesis of creation and annihilation of matter, whether for photons or electrons. Moreover, the clear similarity between this process and the proton-antiproton reaction suggests that the alternative interpretation in terms of real massless particles in inexhaustible supply also has the potential for explaining interactions in the GeV range and beyond. In so doing, we are simply reverting to the ancient proposition of complete elemental balance in all physical transformations by virtue of the assumed indestructibility of the fundamental building blocks of nature.

**III. Correlation of Photon Properties with an Electron-Positron Composition**

A. Intrinsic Properties

The interpretation of positronium decay as an emission process involving different states of the same physical system has been seen to suggest that the photon itself is a compound of a single electron and positron. It is therefore interesting to compare the properties expected for such an $e^+e^-$ structure with those known experimentally for the photon. To begin with, it can be noted that a system containing two fermions in a highly bound state would be expected to obey the Bose-Einstein statistics observed for photons. The spin of the combined system must be integral, just as for positronium in any of its hydrogenic states. Whether a system consisting of an even number of fermions behaves as a boson or not is known to depend on the strength of the interactions holding the individual particles together [7]. The $^3$He isotope, for example, is fermionic and non-superconducting, but combining it with another fermion (the neutron) produces $^4$He, which behaves as a boson.

Otherwise, what we know of photons is that they have zero rest mass and no charge, the latter property being clearly consistent with an electron-positron composition. The fact that photons of a given energy are characterized by a definite frequency and wavelength does not distinguish them from other particles, as emphasized by the de Broglie relation [8], $p= h/\lambda$, and the Bohr frequency law [5], $E=h\nu$, as demonstrated explicitly for electrons by Davisson and Germer [9]. For photons there is the additional feature of oscillating electric and magnetic fields being involved explicitly in the wave motion. Especially for optical photons, the frequency of the oscillations is too large to enable a direct measurement of the individual electric or magnetic fields [10], however. The oscillating properties of photons are actually deduced from theoretical considerations, namely the solution of Maxwell's classical equations of electromagnetism [11]. In quantum mechanics photons have traditionally been treated as oscillating systems [12], without giving a detailed description of the internal structure that is ultimately responsible for such characteristics. All that can be said in the present context is that an $e^+e^-$ composition for the photon is at least consistent with these electromagnetic phenomena. The dipolar nature of such a binary system meshes qualitatively with the photon's capacity for interacting with charged particles, especially when one imagines the photon to be in relative motion with respect to the latter. One would have to have much more detailed information concerning the wave function of the $e^+e^-$ system in a given state of translation to make more specific comparisons with real photons. Similarly, since the speed of the photons is a consequence of their zero rest mass, this is again at least a conceivable property for a system with such a dipolar composition, one whose verification again would require a more quantitative theoretical treatment.

The polarization of light has been one of its most intriguing properties. It has been interpreted by Wigner [13] to result from the fact that the photon possesses non-zero angular momentum **J**. The "twoness" of the photon's polarization is thereby explained as a relativistic requirement, according to which a particle moving with the speed of light must have **J** oriented either parallel or antiparallel to its line of motion. Quantum mechanically this means that only $M_J = \pm 1$ is allowed, despite the requirement of symmetry that components with $M_J < |\mathbf{J}|$ also exist. Circularly polarized light corresponds to an eigenfunction of $J_z$, while plane-polarized implies a 50-50 mixture of both allowed $M_J$ values and elliptically polarized light is any combination in between, all of which is consistent with the existence of an effective two-fold degeneracy. Careful experiments [14-15] have demonstrated that the magnitude of a circularly polarized photon's spin component is h, corresponding to $|\mathbf{J}|= 1$, which is consistent with the Wigner interpretation [13] but also with a possible $e^+ e^-$ constitution for the photon itself.

A related question arises in the present context, however, namely what value of J should be attributed to a massless photon, *i.e.* with E=p=0. Since **L** = **r** x **p**, the most obvious answer is **L** + **S** = 0. There is no contradiction in this conclusion, however, because the photons investigated in the above experiments invariably have non-zero energy. It should be recalled that photons produced in the most common emission processes (E1) involve a change of $\Delta J = \pm 1$ in the associated transition. Conservation of angular momentum requires that a similar change be observed for the photon, and thus J = 0 is possible on this basis for the photon in its initial state, since at the end of the emission process it is found to have J = 1. It will also be seen that the same argument indicates that the parity of the initial and final photon states must be opposite (again when an E1 mechanism is involved), but as long as there is no clear choice dictated for the parity of a massless photon, there is also no contradiction on this basis. It should be noted that the creation and annihilation of matter hypothesis has the emission photon arising from nothing, to which one has difficulty ascribing either angular momentum or parity. On this basis there is no choice but to assume an E1 photon as $1^-$. Since a bound $e^+e^-$ state is distinct from nothing, the same argument cannot be made in the present interpretation, but because there can never be an absolute determination of parity [16] anyway, this state of affairs cannot rule out the massless photon hypothesis under discussion.

Altogether it should be recalled that despite intense investigation over centuries, going back at least to the work of Newton [17] and Huygens [18], there is very little consensus about the structure of the photon itself, or indeed whether it has any internal structure at all. Einstein remarked [19] in 1951 that, despite his best efforts of the preceding half-century, he did not feel that he had come any closer to answering the question of what a light quantum is. He went on to say that apparently many people [20] did think they understood the matter, but that they were only deceiving themselves. At the very least his comments would seem to allow considerable latitude for further research into this question.

B. Compton and Raman Effects and Bremsstrahlung

The Compton effect involves collisions between x-ray photons and weakly bound electrons and can also be interpreted [21] in a very straightforward manner using conventional energy and momentum conservation arguments in conjunction with the Bohr frequency and de Broglie relations. In this experiment a photon with a given energy is scattered off an (essentially free) electron and another photon is observed after the collision with lower energy and momentum than the first. Whether the same (x-ray) photon is involved before and after the collision is an interesting question. In the present interpretation consistency requires that the initial photon give up all its energy and thereby reach a massless state, since only then can the space-time requirements for this interaction be fulfilled. Another initially massless photon takes up the energy left over from the electron collision and appears as an x-ray photon, generally moving in a different direction than the first. To an observer the resulting process is clearly indistinguishable from one in which the incoming photon is annihilated and the outgoing one is created.

The Raman effect [22,23] has similarities with the Compton effect, but involves inelastic scattering of visible light off molecular systems. If the initial frequency is ν, it is found that photons emerge at right angles to the incident radiation with frequencies ν ± ν', where ν' is a characteristic infrared frequency small compared to ν. Again there is the question of whether to interpret the incoming and outgoing photons as one and the same, only with changed energy, but the problematics are clearly the same in this regard as for the Compton effect. In both cases it is clear that a particle interpretation for the electromagnetic radiation allows for a quantitative description of these phenomena.

Finally, it is pertinent to consider the Bremstrahlung phenomenon in this connection as well. In this case an x-ray photon is produced with energy essentially equal to the decrease in an electron's kinetic energy caused by its collision with a heavy nucleus. The process can thus also be thought of as an interaction in which a massless photon picks up energy, similarly as in the emission processes discussed earlier.

C. Production of Particle-Antiparticle Pairs from Photon Collisions

The reverse process to positronium decay, in which an electron and positron are produced with the aid of high-energy photons, also needs to be considered in the present context. The assumption of an $e^+e^-$ structure for each photon is obviously consistent with this result, but a few details require special attention. When a photon with an energy equal to $2\ m_ec^2$ collides with a massless (E=0) photon, no electrons can be produced unless a heavy nucleus is also present. By contrast, if two photons collide head-on, and each has $m_ec^2$ energy, electron production is possible in free space. The distinction can be understood with reference to the theory of special relativity.

A collision between such a massless photon and one with $E = 2m_ec^2$ is characterized by a total momentum of $p = E/c = 2m_ec$. If one of the photons were to dissociate into its elements $e^+$ and $e^-$, all the available energy would be used up for this purpose, so that the translational energy of the two electrons produced would have to be null. The latter condition makes conservation of linear momentum in such a process impossible, however. By contrast, if both photons have E=$m_e c^2$ and collide head-on so that the momentum sum $\Sigma\, p_i = 0$, it follows that the electron and positron can be set free, but must remain at rest in the original inertial system (the one in which the photons are initially observed to have equal energy). The latter interaction is seen to be simply the reverse of the positronium decay process, or more precisely the reverse of the interaction of a free electron and positron which are initially at rest in a given coordinate system (note that on the basis of Fig.2 an E=0 photon is assumed to actually provide the electron and positron observed at the conclusion of this experiment).

More generally, it needs to be recalled that for a given energy E the momentum of the photon (E/c) is always greater than for any particle with rest mass $m_A > 0$, for which $p_A = (E^2/c^2 - m_A^2 c^4)^{1/2}$. This fact prevents a single photon of any energy from causing a zero-energy photon to dissociate, because no matter what energy might be transferred, there is a disparity in the corresponding photon momentum lost and that which could theoretically be given to each of the electron products. The presence of a third body has the potential of removing this restriction, as is well known, but the point to emphasize in the present discussion is that the same result is found whether free space is thought to be involved, as foreseen in the creation-annihilation hypothesis, or if a massless *but existing* photon of $e^+e^-$ structure is assumed instead.

To make this point more clearly, it is interesting to consider the effect of relative motion of the observer on the outcome of such experiments. The relativistic Doppler effect [24] tells us that the energy (frequency) of the photons in the above examples is dependent on the relative speed of the inertial system from which these quantities are measured. There is a clear exception to this rule, however, namely if the energy of the photon is zero in one inertial system, it must remain zero in any other. Thus it is not possible to make the transition between the above two cases simply by changing the relative speed of the observer. As noted elsewhere [6], a massless photon corresponds to a null vector in Minkowski space [25], and as such is unchanged by any Lorentz transformation. At the same time, a photon with non-zero mass can have its energy changed to any conceivable positive value other than zero by virtue of such a transformation. The consequences of these relationships are crucial in the present case, with electron-positron production in "free space" occurring only if both photons have non-zero energy, just as is observed experimentally.

With much higher energies it is also possible to generate proton-antiproton pairs [3], again as predicted by the Dirac theory [1]. It is clear that this result cannot be entirely explained by assuming an $e^+e^-$ structure for the photon. Nonetheless, it cannot be said that such observations are inconsistent with what has been assumed so far. Rather they force an additional assumption, namely that other types of massless particle-antiparticle binaries exist as well. There is a natural tendency to avoid introducing new types of particles into any theoretical framework, however, and at the very least one hopes to keep their number to an absolute minimum.

As long as its rest mass were exactly zero, the mechanical properties already mentioned for $e^+e^-$, such as $v = 0$, $\lambda = \infty$ and the like, could also apply to $p^+p^-$ or related entities. One can only speculate that a $p^+p^-$ system of zero rest mass will exhibit different properties under translation than do the corresponding $e^+e^-$ species. Clearly, the dissociation energy of $p^+p^-$ must be 1836 times greater than for $e^+e^-$, which condition already constitutes a major distinction. By the same token, the fact that neutron decay produces neutrinos [26], whose rest masses are already close to or equal to zero, implies that there must be $\nu\nu$ binaries as well, with extremely small to vanishing dissociation energies. The real challenge presented by these observations is to construct a quantitative theory, requiring as input at most such quantities as the rest mass, charge and perhaps magnetic moment (q/m) of the interacting species, *which leads to binding energies of the above particle-antiparticle pairs which are exactly equal to $2c^2$ times the rest mass of each of the respective constituents.*

Since the charge-to-mass ratio is much smaller for the proton than the electron, it seems clear that a $p^+p^-$ binary would show much weaker electromagnetic effects than its $e^+e^-$ counterpart. On this basis, it seems plausible that the traditional properties of a photon, *i.e.* an oscillating electromagnetic field that is involved even in low-energy emission and absorption processes, are *exhibited exclusively by the electron-positron massless binary systems*. Nonetheless, the statistical arguments given in conjunction with the discussion of blackbody radiation [6,27] would seem to speak equally well for a high density of other (bosonic) systems of zero rest mass. At least one knows that protons and antiprotons can be produced together at any location where the appropriate energy and momentum conditions can be fulfilled.

It should also be noted in this connection, however, that it is not necessary to assume that every particle forms a massless binary system with its antiparticle. Since the neutron and antineutron are known to decay into three stable particles each after a relatively short lifetime, for example, it would be entirely consistent with the present view to assume that only massless particle-antiparticle combinations of their respective (stable) decay products are formed when these two systems interact. Moreover, once the hypothesis of creation and annihilation of matter is questioned, it becomes necessary to come up with another explanation for the appearance of an electron and antineutrino at the conclusion of the neutron decay process. At the very least, it would seem justified to pursue such an alternative view by attempting to characterize (short-range) forces that are capable of binding an electron within the nuclear volume. A convenient place to begin such a search is the positronium decay process discussed in Sect. II. If a reasonable concretization of the massless $e^+e^-$ state indicated at the bottom of Fig. 1 can be obtained, the discussion of the present section indicates that such a system could then be realistically identified with the photon itself.

**IV. Variational Theory for Particle-Antiparticle Interactions: Concretization of the Elektroweak Interaction**

In the preceding discussion it has been suggested that there is merit in considering the decay of positronium as an interaction in which an electron and positron are so strongly attracted to one another that the resulting binding energy is exactly equal to the sum of their rest masses, *i.e.* $2m_ec^2$ or 1.022 MeV. A survey of the key experiments in modern physics has shown how such a development would fit into the theory of elementary particles and the energetics of their reactions. In the present section we will focus on the goal of finding a suitable potential which is capable of producing such a relatively large binding energy, while at the same time giving consideration to the possibility that the solution to this problem may have relevance for other types of interactions as well, particularly those involved in the study of nuclear and elementary particle physics. This includes especially the weak interaction, which is assumed to be responsible for neutron decay and related processes.

It is worthwhile to note at this juncture that the interaction of an electron with its antiparticle is one of the very few processes observed experimentally for which no potential has been sought to explain its occurrence. The usual approach is simply to calculate the amount of energy produced in the process as $2m_ec^2$ from Einstein's mass/energy equivalence relation, with only the relatively vague implication that the forces involved must be of short range and not just Coulomb in nature. We know from the study of nuclear physics that mass defects for such systems merely allow one to calculate the exothermicity of a given reaction, however. It is always understood that there is a well defined, if not always thoroughly understood, force at work which causes nucleons to interact so strongly with one another as to produce the observed mass defect. It is therefore consistent to assume that a specific force is also responsible for the electron-positron and other particle-antiparticle interactions, even though the energy loss itself in these cases is again easy to calculate without knowledge of the exact nature of these interactions.

A. A Short Range Potential

A natural point at which to begin this investigation is with the nature of the potential that might be capable of producing the strong attraction between an electron and a positron, although careful consideration must also be given to the manner in which the kinetic energy is treated as well. It is clear from the outset that this must be a distinctly relativistic problem, because the rest mass of the combined $e^+e^-$ system is assumed to be much lower than the sum of those of the separated particles. Both the Schrödinger non-relativistic [28] and Dirac relativistic [29] treatments of positronium tell us that the lowest possible state for this system is analogous to the 1s state of the hydrogen atom (Fig. 1), in which case the primary interaction is Coulombic, exclusively so in the non-relativistic treatment and almost exclusively in the relativistic.

A two-component reduction of the Dirac equation leads to the characterization of a number of perturbative terms that are basically magnetic in nature. The most commonly employed such approximation is that of Breit-Pauli theory [30-32], including the Breit interaction. The perturbations are of the order $\alpha^2/2 \cong 10^{-5}$ hartree ($\alpha = e^2/hc = 137.036^{-1}$, the fine-structure constant, hartree = 27.21 eV), and include the spin-orbit (same- and other-orbit), spin-spin, orbit-orbit and Darwin terms, as well as the mass-velocity correction to the non-relativistic kinetic energy [32]. These terms vary as $Z^3$ or $Z^4$ (spin-same-orbit) for atoms with nuclear charge Z. For positronium as well as the hydrogen atom they remain quite small, however, and their effects are observed only as fine structure in spectroscopic studies. The potential terms all vary as $r^{-3}$ and thus have relatively short ranges compared to the Coulomb interaction. This point bears further consideration, however, since binding energies of 1.0 MeV and higher are otherwise known only for nuclei, in which case there is clear evidence that short-range forces are involved to a high degree.

We can represent the presumed interaction schematically by plotting the total energy as a function of the average distance r between an electron and positron (Fig. 3). The 1s state of positronium can be thought of as corresponding to a minimum in total energy occurring at r=2.0 bohr = 1.016 Å, i.e. roughly double the corresponding value for the H atom by virtue of the smaller reduced mass of the $e^+e^-$ system. Toward larger separations the energy gradually increases to zero, i.e. the energy of the separated particles. The attractive Coulomb potential varies as $r^{-1}$, while the kinetic energy in this region varies as $p^2 \sim r^{-2}$, from which it follows that the total energy itself at first *decreases* as the particles approach one another from a large distance. At the location of the energy minimum the shorter range of the kinetic energy term becomes the dominant factor, which explains why the total energy thereafter increases rapidly toward still shorter distances. It will be recognized that these arguments are very close to those used by Bohr [5] in arriving at his theory of hydrogenic atoms in 1913, simple but effective.

The binding energy at the latter $e^+e^-$ minimum is only 6.8 eV, which is quite small compared to the 1.022 MeV given off when positronium decays from the corresponding (1s) state. The possibility we wish to explore in this work is *whether a second energy minimum might not occur at a much smaller electron-positron separation*. It can be imagined, for example, that at some point the total energy stops increasing toward shorter distances because an attractive short-range potential term begins to overcome the effects of increasing kinetic energy in this region. Such a potential term would have to vary at a higher inverse power of r than either of the other two terms in the non-relativistic electrostatic Hamiltonian, and would have to be relatively unimportant in the region of the first hydrogenic energy minimum. At the same time, it is evident that some effect with an even shorter range eventually must take over and cause the energy to increase (Fig. 3) once more toward even smaller distances after the assumed 1.022 MeV absolute minimum is attained. It is also clear that such a second energy minimum must be *totally absent in the corresponding hydrogen atom treatment.* Finally, it is only consistent to assume that an analogous double-minimum curve exists for the proton-antiproton system, but with a binding energy that is 1836 times larger, i.e. 1.876 GeV for its short-range minimum. At least one knows that this much energy is given off when the proton and antiproton interact, whereas no comparable loss of energy is observed for the combination of a proton and an electron.

Concentrating on the comparison of $e^+e^-$ with $p^+e^-$, the obvious question is how can the differences in the properties of the proton and positron lead to such an enormous distinction in their respective attractions to the electron. The traditional view embodied in the Schrödinger [28] and Dirac [29] equations for one-electron atoms holds that the large difference in mass of the two positively charged particles plays only a minor role in this connection, simply affecting the reduced mass of the electron. The magnetic moments (which have the charge-to-mass ratios as a factor) of $e^+$ and $p^+$ differ by a far greater amount because of the difference in their rest masses, but this distinction is found to be of only minor importance in the Dirac equation treatment, in which effects such as the spin-orbit and spin-spin interactions which depend on this quantity are accounted for explicitly.

Yet one knows from the outset that if there is indeed a much lower-lying state of the $e^+e^-$ system than the familiar 1s species, it cannot be found among the solutions of the hydrogen atom Dirac or Schrödinger equations. To progress further in this regard it is necessary to do something differently. Especially since the effect that might cause such a novel tight-binding $e^+e^-$ state seems almost certainly short-range in nature (Fig. 3), there is reason to give closer consideration to the above magnetic-type interactions. As noted previously, there are numerous Breit-Pauli terms which fall in this category, varying as the inverse cube of the distance between interacting particles. They are all of order $\alpha^2/2 \cong 10^{-5}$ hartree for typical atomic electron-nucleus separations, so this characteristic fulfills another requirement from Fig. 3, namely that such a short-range effect be relatively insignificant at these distances.

All the above Breit-Pauli terms [32] depend on the product of the magnetic moments (or charge-to-mass ratios) of the interacting particles. For the spin-other-orbit, spin-spin and orbit-orbit terms each of these quantities appears once in the corresponding product. Thus these terms are weighted by a factor of 1836 (the ratio of the rest masses of proton and positron) larger for $e^+e^-$ than for $p^+e^-$. For the spin-same-orbit and Darwin terms the distinction is less important because in these cases the square of the mass of one of the constituent particles is involved rather than the product of both. As a result, there is an extra factor of two for these interactions for e$^+$ e$^-$ than for the hydrogen atom by virtue of the fact that the square of the charge-to-mass ratio of the proton is negligible compared to that of the positron. At atomic distances these distinctions are still relatively unimportant because in this range the Coulomb interaction dominates, but it is not difficult to imagine the situation could be far different at shorter range.

Before considering this possibility further, however, it is well to note that the Breit-Pauli Hamiltonian terms also have some undesirable properties which make them unsuitable for a variational calculation (i.e. one in which the charge distributions of the particles involved are allowed to assume optimal forms so as to minimize the total energy). Since these terms vary as $r^{-3}$, there is nothing to keep the total energy from decreasing beyond any limit. They do not therefore provide a possibility of a second energy minimum of the type indicated in Fig. 3, rather only the attractive branch to the long-distance side of it. This fact has long deterred giving serious consideration to the Breit-Pauli terms as having any truly dominant role to play in quantum mechanical calculations. There are other higher-order terms in a power-series expansion of the Dirac equation that need to be considered to properly understand their role in determining atomic fine structure, however. Specifically, the next terms in such expansions are of the order of $\alpha^4 r^{-4}$, and these higher-order species prevent variational collapse in Dirac equation solutions that would otherwise occur if only the $\alpha^2 r^{-3}$ terms were included.

A similar situation exists for the Lorentz force Hamiltonian in classical electromagnetic theory. There one has a term of the form $|\mathbf{p} + e\,\mathbf{A}/c|^2$, where $\mathbf{A}$ is the vector potential. The cross term involving $\mathbf{p}\cdot\mathbf{A}$ is typically [33] also of the form $\alpha^2 r^{-3}$ ($c = \alpha^{-1}$ in atomic units) for the interaction of a charged particle with an electromagnetic field, but it is damped at short range by the $|\mathbf{A}|^2$ term, which varies as $\alpha^4 r^{-4}$. The fact that such repulsive terms *are of even shorter range and higher order in $\alpha^2$* than their Breit-Pauli counterparts nonetheless opens up an interesting possibility. Including such terms might not keep the total energy in Fig. 3 from turning downward at short distances because of the attractive $\alpha^2 r^{-3}$ interactions, but they would insure that this trend not continue indefinitely toward still smaller interparticle separations, with the result that the proposed non-hydrogenic second energy minimum could be formed. If one simply assumes a potential which is the difference of these two short-range terms, $-\alpha^2 r^{-3} + \alpha^4 r^{-4}$, it is easily shown that it possesses a minimum near $r \cong \alpha^2$. Such a distance corresponds to roughly $10^{-5}$ bohr, which is a typical separation for bound nucleons ($r=\alpha^2/2$ is normally given for the range of the nuclear force, for example). The possible connection between an e$^+$e$^-$ tight-binding state and known short-range interactions is thus reinforced by these considerations as well.

There are basically three arguments for pursuing this line of reasoning further. First, the theory of quantum electrodynamics has a very precisely defined range of applicability. Despite its ability to make extremely accurate predictions for interactions involving electrons and photons in an atomic environment, it is generally accepted that the theory in its established form is not capable [34] of describing high-energy interactions such as those involved in nuclear binding. In more recent times it has been possible to obtain a unification of the electromagnetic and weak interactions [35-38], however, which development at least suggests that a Hamiltonian exists which when employed in a Schrödinger equation or equivalent theoretical treatment is capable of describing both Coulomb and shorter-range interactions on an equal basis. Any indication as to how *the computational framework of quantum electrodynamics could be further adapted so as to become more quantitatively applicable to the description of short -range forces* therefore merits serious consideration.

Secondly, the analysis of the positronium decay process has provided a basis for associating the assumed $e^+e^-$ tight-binding state with the photon itself. Given the prominent role of the photon in quantum electrodynamics, it seems plausible that its internal structure would also have an important relationship to the electromagnetic force. Finally, examination of the multiplet structures and other properties exhibited by nuclei has already led to the conclusion [39-40] in the nuclear shell model that spin-orbit or related terms [41] are almost certainly involved in this type of high-energy interaction . Taken together these observations suggest that a solution to the proposed problem may lie in an adaptation of the Dirac equation, with suitable radiative corrections, which does not detract from the reliability of the original theory's predictions for quantum electrodynamical phenomena, but which at the same time enables an accurate treatment of interactions of much shorter range and higher binding energy.

B. Kinetic Energy Considerations

As remarked above it is not satisfactory to use a non-relativistic form for the kinetic energy of the positron and electron if a large (MeV) binding energy is assumed. In this respect, the problem is significantly different from the conventional treatment of nuclear binding, because there the kinetic energies of the nucleons are still relatively small compared to the energy equivalent of their rest masses. In 1905 Einstein showed [4] on the basis of the special relativity theory that the non-relativistic $p^2/2m_o$ term is actually just a term in the power series of the square-root quantity $(p^2c^2 + m_o^2c^4)^{1/2} - m_oc^2$. The Breit-Pauli approximation includes [32] a term of order $p^4$ to account for relativistic kinetic energy contributions, but just as for the corresponding potential terms, it is known that this correction leads to variational collapse when the electronic charge distribution is allowed to vary freely so as to minimize the energy. It is possible to circumvent this difficulty [42], however, by including the Einstein square-root expression directly in the Hamiltonian instead of relying on a truncated power-series expansion for it. The inconvenience of employing a square-root operator can be dealt with for general atomic and molecular systems by means of a standard matrix procedure. Each particle has its own kinetic energy and thus the square-root terms are treated as one-electron operators, exactly as their non-relativistic $p_i^2/2\,m_i$ counterparts in conventional quantum mechanical treatments.

This choice makes it much more difficult to separate the total kinetic energy into internal and center-of-mass components, in contrast to what is done in non-relativistic theory. This state of affairs is also found in relativistic quantum field theory and thus should not be a source of concern, however. It causes the traditional internal properties of non-relativistic theory to be dependent on the state of translational motion, which characteristic is necessary to ensure that the predictions of the theory regarding energies and lifetimes be consistent with the Lorentz transformation. The inclusion of momentum-dependent short-range interactions in the Hamiltonian has a similar effect, but this also requires the introduction of a simple transformation that can separate out the motion of the center of mass. In the following section it will be shown that this property of the assumed potential provides a simple explanation as to why an electron and positron can be as strongly bound together as indicated in Fig. 1, with 1.022 MeV exothermicity, whereas a proton and electron can only have a maximum binding energy of a few electron volts.

Under the circumstances the most convenient procedure in designing the present theoretical description is to simply work with the original Cartesian coordinates of each particle. The quantum electrodynamical treatment of the positron uses a similar approach [43,44] in evaluating higher-order effects, including transition probabilities for the decay into photons from various positronium electronic states, ultimately imposing an additional condition of $\mathbf{P} = \Sigma\mathbf{p_i} = 0$. In such a two-particle application this means that $\mathbf{p_i} = -\mathbf{p_2} = \mathbf{p}$. No comparable transformation is employed for the spin coordinates of both particles, however. Radiative corrections to the Dirac equation results can also be computed by employing the same condition for the center-of-mass momentum [43-44]. When more than two particles are involved complications arise in attempting to generalize this procedure, however, particularly in the relativistic treatment of the kinetic energy. In the transformed coordinate system, one particle is effectively singled out as a reference for the internal coordinates, so that $x'_i = x_i - x_1$ is now employed instead of $x_i$, for example (with $x_1$ itself being retained in the new basis). As a result the expression for the conjugate momentum of $x_1$ is $\mathbf{p_1} = -\sum_{i \neq 1}{}' \mathbf{p'_i}$, *i.e.* a sum of internal momenta rather than a single such quantity as in the two-particle case.

C. Suggested Damped Form for the Breit-Pauli Potential Terms

A prerequisite for constructing a Schrödinger equation to investigate high energy processes is the use of a potential which is suitably bounded. It has been shown in Sect. IV.A that the assumed tight-binding $e^+e^-$ state most likely is the result of a short-range interaction which is strongly attractive over a given region of interparticle separation but even more strongly repulsive at still smaller distances. The Breit-Pauli approximation [30-32] employs attractive terms fitting this description but lacks corresponding shorter-range effects that would produce the desired second minimum in the $e^+e^-$ total energy curve sketched in Fig. 3. We have seen how the Breit-Pauli kinetic energy term (including the mass-velocity correction) can be replaced by the Einstein free-particle square-root expression to avoid variational collapse without giving up the advantages of having a reliable approximation to the Dirac equation results for hydrogenic atoms. It remains to find a similar means of dealing with the Breit - Pauli potential terms.

If we look upon the spin-orbit and related interactions as terms in a power series, it seems reasonable to search for a closed expression that approaches this result in the low-energy relativistic regime encountered in calculations of atoms with moderately heavy nuclei - something analogous to the Einstein relativistic kinetic energy, in other words. At least one knows from the form of the Lorentz electromagnetic force that the next higher-order terms after those of $\alpha^2 r^{-3}$ spin-orbit type vary as $\alpha^4 r^{-4}$. Simply adding such terms to the Hamiltonian has several disadvantages, however. It falls short of the goal of replacing the Breit-Pauli interactions with closed expressions which themselves correspond to infinite-order power series. In addition, it is difficult to carry out computations with an operator varying as $r^{-4}$ because not all integrals which would be required in a variational treatment are finite, in particular not those involving only s-type basis functions.

The form of the desired potential (Fig. 3) is reminiscent of that observed in nuclear scattering, as mentioned earlier, and this realization suggests the following alternative, however. The $r^{-3}$ and $r^{-4}$ terms already discussed can be grouped together as $\alpha^2 r^{-3} (1- \alpha^2 r^{-1} + ...)$. The terms in parentheses are the beginning of a power series expansion of the function $\exp(-\alpha^2 r^{-1})$, which in turn is similar to that appearing in the Yukawa potential [45] of nuclear theory, except that in the latter case the exponential argument is proportional to r rather than its inverse as in the present case. By multiplying the Breit-Pauli $\alpha^2$ terms with such an exponential function, we have a potential that is capable of producing the second minimum for $e^+e^-$ in Fig. 3, while at the same time retaining the correct behavior needed at relatively large interparticle separations to properly describe the conventional positronium (hydrogenic atom) states. Since the damping effects produced by the exponential factor are relativistic in nature, it seems somewhat more likely that the corresponding argument is a function of the momentum of the particles rather than the distance between them, *i.e.* of the form $\exp(-\alpha^2 p)$, with $r^{-1} \sim p = |\mathbf{p}|$. This choice has computational advantages as well, because it means working with individual quantum mechanical operators that depend on the coordinates of a single particle rather than two. From the Lorentz classical Hamiltonian we can also anticipate that a given particle's momentum p is multiplied by its charge-to-mass ratio $q/m_0$. Finally, to ensure the desired binding energy for a given particle-antiparticle system it is convenient to introduce a free parameter A as an additional factor in the exponential argument which, on the basis of the above arguments, should turn out to be of the order of unity when atomic units are employed. he negative of the resulting potential is plotted in Fig. 4 (atomic units are employed throughout) as a function of the reciprocal interparticle distance $r^{-1}$. For this purpose we use the approximate representation $V(r)= -\alpha^2 r^{-3} \exp(-\alpha^2 r^{-1})$. For high particle velocities it can be assumed that the kinetic energy varies as pc in the range of interest, which can therefore be represented in an analogous manner as $\alpha^{-1} r^{-1}$ in atomic units, *i.e.* as a straight line. This diagram is useful in analyzing how binding can be achieved with such an exponentially damped potential over a very narrow range, consistent with the total energy curve shown in Fig. 3. For small momenta typical of electrons in atoms, the kinetic energy far outweighs the short-range potential because of the factor of $\alpha^2$ in the latter expression. Coulomb effects are omitted from consideration for the time being.

The absolute value of the potential increases as the cube of the momentum (or reciprocal distance) while the kinetic energy changes in a nearly linear manner, so it can be imagined that the two quantities eventually become equal at some point and binding becomes possible. The exponential damping should become noticeable in the same region, however, causing the above term not to increase as quickly as before and finally reach a maximum. At the same time, since the kinetic energy must continue to increase linearly with p, thus eventually producing a second crossing with the negative of the potential (Fig. 4). The area in which the negative of the potential exceeds the kinetic energy corresponds to a very small range of r, but the amount of binding with which it is associated can still be quite large. For example, at r= $\alpha^2$ the *undamped* Breit-Pauli potential is of the order $\alpha^{-4}$ hartree, compared to the kinetic energy's order of $\alpha^{-3}$ hartree.

Since the assumed binding energy for the $e^+ e^-$ system is 1.022 MeV or $2\alpha^{-2}$ hartree, it is clear that an enormous cancellation must occur because of the damping of the potential in order to obtain physically acceptable results. This state of affairs is probably the strongest argument for employing an exponential damping to effect such a large degree of binding over a narrow range of interparticle separation. The fact that the Coulomb energy is also of order $\alpha^{-2}$ hartree for r = $\alpha^2$ suggests that it is not possible to ignore this effect either, however, despite its relatively long-range character. Nonetheless, the predominant feature in the tight-binding scenario given above is clearly the delicate cancellation at small interparticle separations between the exponentially damped Breit-Pauli terms and the relativistic kinetic energy.

**V. Key Characteristics of the Proposed Theory**

A. Scaling Properties of the Breit-Pauli Hamiltonian

One of the requirements for the Hamiltonian and related Schrödinger equation under discussion is that they produce maximum binding energies for particle-antiparticle systems of $2Mc^2$, consistent with the Einstein mass-energy equivalence relation. Instead of assuming that annihilation occurs and the total mass of the particles simply appears as the equivalent amount of energy, one postulates that a Hamiltonian exists which has the required energy as its *minimal* eigenvalue. It is well known that the Schrödinger and Dirac equations for purely electrostatic potentials both have the property that their binding energies are proportional to the reduced mass of the electron in hydrogenic atoms, and this result is easily generalized for systems containing other charged particles such as muons or antiprotons. More interesting in the present context is the fact that the proportionality between energy and mass also holds when the various Breit-Pauli relativistic corrections are added to the Hamiltonian.

To show this let us assume that a solution to the Schrödinger equation is known for a particle-antiparticle pair with charge q and rest mass $m_0$. Furthermore, its lowest energy eigenvalue is taken to be $-2m_0c^2 = -2m_0\alpha^{-2}$ (in atomic units), corresponding to the eigenfunction $\Psi(r)$. The Hamiltonian itself consists of a series of kinetic and potential energy operators of the type discusssed earlier, including exponential damping factors $F(p,q,m_0)$ :

$$H(p, r, q, m_0) = (p^2\alpha^{-2} + m_0^2\alpha^{-4})^{1/2} - m_0 \alpha^{-2}$$
$$- q^2 r^{-1} - q^2 m_0^{-2}\alpha^2 r^{-3} F(p, q, m_0) \qquad (7)$$

If the coordinates are scaled so that
$$p' = M_0 m_0^{-1} p \text{ and } r' = M_0^{-1} m_0 r, \qquad (8)$$
the original Hamiltonian becomes:

$$H(p, r, q, m_0) = M_0^{-1} m_0 \{(p'^2 \alpha^{-2} + M_0^2\alpha^{-4})^{1/2} - M_0\alpha^{-2}$$
$$-q^2 r'^{-1} - q^2 M_0^{-2}\alpha^2 r'^{-3} F(p', q, M_0)\}$$
$$= M_0^{-1} m_0 H(p', r' q, M_0), \qquad (9)$$

provided $F(p, q, m_0) = F(p', q, M_0)$. The corresponding Schrödinger equation in the scaled coordinate system thus becomes:

$$H(p', r', q, M_0) \psi(r) = -2 M_0\alpha^{-2} \Psi(r) \qquad (10)$$

*i.e.* by multiplying both sides of the Schrödinger equation for the original Hamiltonian by $M_0 m_0^-$

[1]. As a result it is seen that $\psi(r)$, or the function $\psi'(r)$ obtained by making the corresponding coordinate substitution for it, is an eigenfunction of the analogous Hamiltonian for a particle-antiparticle system of the same charge q as before, but with rest mass $M_0$ instead of $m_0$. Its energy eigenvalue is $-2M_0\alpha^{-2}$, exactly as required by the mass-energy equivalence relation.

Moreover, this result is quite general, since it is easily seen that the above scaling procedure has the effect of producing an entire spectrum of Schrödinger equation eigenvalues which differ by a factor of $M_0 m_0^{-1}$ from those obtained for the original particle-antiparticle system. Furthermore, by choosing the argument of $F(p,q,m_0)$ to contain the ratio $p/m_0$ as a factor, as suggested by the form of the Lorentz electromagnetic force Hamiltonian discussed in Sect. IV, the requirement that this damping factor be unaffected by such a coordinate scaling is clearly fulfilled. The Breit-Pauli interactions also contain angular orbital momentum terms not included explicitly in the above Hamiltonian, but these are unaffected by the above scaling procedure because they either involve only products of r and p, or in the case of the spin interactions, are completely independent of spatial coordinates. It is thus demonstrated that the desired proportionality between binding energy and rest mass of the constituents of a particle-antiparticle binary system holds for the Breit-Pauli interaction as long as the charge q of the individual particles remains the same. This is clearly the case in comparing the $e^+e^-$ system to $p^+p^-$, so the original objective sought at the beginning of Sect. IV is guaranteed by the use of such a Hamiltonian.

This result also tells us that the proportionality constant A in the exponential damping factor $F(p,q,m_0) = \exp[-A\alpha^2|(q/m_0)p|]$ must be the same for proton-antiproton interactions as between electron and positron. Alternatively, one might have assumed a different constant for the electron than for the proton, in which case one would have had to adjust the inverse mass dependence of the present exponential argument in order to obtain the desired scaling property. From the point of view of economy of assumptions and relation to established theoretical models, the former arrangement is clearly superior. The units for the constant A are left somewhat open by the choice of working in the atomic unit system. The exponential argument as a whole must be dimensionless, but the fine structure constant ($\alpha$ can be looked upon as either $e^2/hc$ or simply as $c^{-1}$ (among other possibilities). Since $p/m_0$ has units of velocity one possibility is to look upon one of the $\alpha$ factors as $c^{-1}$ and the other as being dimensionless, in which case A must have units of reciprocal electric charge, so that Aq is also dimensionless. An explicit definition of the exponentially damped Breit-Pauli Hamiltonian discussed above is given in Table 1.

B. Qualitative Explanation for the Distinction in $e^+ e^-$ and Hydrogen Atom Maximal Binding

Energies

In Sect. IV.B it was noted that the usual separation of internal and center-of-mass motion is not possible for a Hamiltonian containing the Einstein relativistic kinetic energy operator and the various Breit-Pauli terms (damped or otherwise). This observation deserves closer examination because it forms the basis for a qualitative understanding of why $e^+ e^-$ (and $p^+p^-$) can have the proposed highly bound ground state while the hydrogen atom has only a relatively small ionizing energy.

A condition of zero translational energy requires that the expectation value of $\Sigma \mathbf{p_i}$ vanish, which for a binary system such as positronium or the hydrogen atom implies that the momenta of each particle are equal and opposite to one another at all times and spatial positions. Since $\mathbf{p} = m\mathbf{v}$, this means that particles of equal mass must always move with equal speeds in opposite directions relative to their midpoint to fulfill the condition of zero translation. The Breit-Pauli interactions are not only short-range but also momentum-dependent (Table 1), and so the only way to obtain a large attraction on the basis of such terms is for the expectation values of $p_i$ and $r_{ij}^3$ to *both* be large for a given probability distribution. From the above argument, however, it is clear that a positron and an electron, with their equal rest masses, can stay in close proximity to one another (i.e. with $\mathbf{v_1} = -\mathbf{v_2}$) while still moving at high velocity, without having net translation for the system as a whole. Hence a high degree of binding can result from this type of short-range interaction under these circumstances.

By contrast, if the masses of the two particles are quite different, as is the case for the proton and electron in the hydrogen atom, they must move with widely different speeds to avoid net translation. Particles whose speeds differ by a factor of 1836 can only stay close to one another while fulfilling this condition *if neither is moving very fast,* from which it can be seen that the possibilities for obtaining a tightly bound state under these circumstances are much less favorable. On this basis, it is easy to imagine how the $e^+ e^-$ system might benefit much more strongly from inclusion of the Breit - Pauli interactions in the Hamiltonian than does the hydrogen atom. In particular, $e^+e^-$ might possess a state of much lower energy than the familiar 1s species that results primarily because of the long-range Coulomb effect, whereas no comparable tightly bound state could be expected for the hydrogen atom.

At the same time the scaling property discussed in Sect. V.A tells us that the range of the $p^+p^-$ interaction must be shorter than for $e^+ e^-$ by a factor equal to the ratio of the respective rest masses of the proton and electron. The $\mathbf{v_1} = -\mathbf{v_2}$ condition for motion without net translation also applies to the $p^+p^-$ system and so one can explain the even larger binding energy of $p^+p^-$ on the same basis. Their larger mass allows the proton and antiproton to approach each other far more closely on the average than the electron and positron while avoiding net translation of either system. The exponential damping factor $F(p,q,m_0)$ plays a crucial role in this distinction, particularly the inverse mass dependence of its argument, since it first begins to effectively negate the influence of the attractive Breit-Pauli terms at much larger separations for the electron and positron than for their heavier counterparts (Table 1).

**VI. Conclusion**

The central issue in the present work is whether material particles can be created or annihilated, passing to or from nothing in the process. Despite the fact that this hypothesis occurs frequently in key interpretations of theoretical physics, it clearly defies direct experimental verification. One can never be certain that the reason some object passes from view is because it has ceased to exist. It can always be that the measuring devices employed to detect it are simply inadequate for this purpose. It is therefore prudent to consider alternative explanations for processes in which one conventionally speaks of the creation and annihilation of matter. The only way this can plausibly be done is to assume instead that there is a strict elemental balance in the corresponding physical transformation, just as is the rule in the theory of chemical and nuclear reactions.

This line of reasoning suggests a new interpretation of the positronium decay process. It is possible to look upon this interaction as a two-photon (in the case of $J = 0$ decay) emission; one simply assumes that what has occurred is a transition between two states of the same ($e^+ e^-$) system. Since one knows that the energy carried away by the photons is exactly equal to the rest mass of the initial system multiplied by $c^2$, it follows that the energy of the final state must vanish, *i.e.* the corresponding mass of the system must be exactly zero. Rather than equate this with the non-existence of the electron and positron constituents of the original positronium atom, however, it is instructive to think of a product of the same composition being formed, one which is so tightly bound that it has exactly zero rest mass. We know from experience in nuclear physics, for example, that a partial loss of mass can be measured when protons and neutrons are bonded to one another. In such cases the mass defect is generally of the order of only one percent of the total, but the theory of special relativity gives no indication that mass changes occurring in other processes must be limited to this range. By analogy it becomes plausible to conclude that electrons and positrons are also not destroyed in positronium decay, but rather that they form an extremely tightly bound diatomic system with the same e+ e- composition as at the beginning of the process.

Since there is already another system of zero rest mass involved in this process, it is possible to make a further unifying assumption, however, namely that the photon and the tightly bound $e^+ e^-$ binary system are one and the same. In this way all the participants in the most common ($J = 0$) positronium decay are taken to have the same $e^+e^-$ constitution (Fig. 2), and the process becomes characterized by complete elemental balance, just as in a conventional chemical or nuclear reaction. At the beginning, two of the $e^+ e^-$ species are (undetectable) E = 0 photons and the other is positronium. As a result of the decay process, the former two particles take on 0.511 MeV of energy each and thus can be detected experimentally, while the positronium species itself undergoes a transition to a massless photon state.

This interpretation has led to several areas of investigation in which to test the $e^+ e^-$ assignment for the internal structure of the photon. First of all, the question can be posed as to why the E = 0 state of the $e^+ e^-$ system is always reached in positronium decay, and not some other state with non-zero translational energy. Consideration of the properties expected from relativity theory for a real system which possesses no mass leads to the conc1usion that in this state the photon is no longer restricted to move at the speed of light. There is no reason that it cannot be at rest with respect to a given observer on this basis. Since there is a clear need for the interaction between the electron and positron to occur in a narrow region of space, this property of massless photons gives them a crucial advantage over their counterparts of non-zero translational energy, which must leave the scene of the interaction at the speed of light. It is therefore far more probable that the original positronium system goes directly to its massless state since it can remain in a fairly localized region of space in this way throughout the course of the reaction. The energy lost must consequently be completely transferred to other particles in the area, which under normal circumstances means that two photons which have E=0 prior to the decay process suddenly appear with equal energy and moving in opposite directions to one another. A similar argument has been employed in a companion paper to explain the nature of photon interactions in conventional radiative absorption processes [6].

A review of the intrinsic properties of a photon has shown them to be quite consistent with such an $e^+ e^-$ dipolar composition. A tightly bound combination of two fermions conforms to the requirement that Bose-Einstein statistics be obeyed. The fact that the photon is electrically neutral and otherwise participates in electromagnetic interactions is also consistent with this structure. The potential for having a binding energy equal to the rest mass equivalent of two electrons clearly allows for the photon's capacity to move with the speed of light. The E=0 state can only correspond to zero angular momentum, but this is seen to be consistent with the fact that photons of non-zero energy produced from their massless counterparts in E1 processes (with a $\Delta J = \pm 1$ selection rule) are measured to have J = 1. Furthermore, evidence from the Compton and Raman effects, Bremsstrahlung and other experiments can also be satisfactorily interpreted in terms of such a definite composition for the photon. Finally, the creation of electrons and positrons from photons is easily understood in terms of this composition, merely involving dissociation of a diatomic system when sufficient energy is supplied and other dynamical conditions are satisfied.

To give further substance to the present structural hypothesis it is necessary to characterize an interaction which can plausibly account for the proposed $e^+$ $e^-$ binding energy of 1.022 MeV, over 150 000 times more than is known for positronium in its lowest state. This line of approach raises another question about the creation and annihilation of matter hypothesis, however. In that view, one simply calculates the energy given off when an electron and positron come together by using the Einstein mass-energy equivalence relation, thereby leaving at least the impression that the forces responsible for this large expenditure of energy are somehow completely described in this manner. Yet one knows from nuclear physics that a mass defect which occurs as a result of a binding process is always attributed to a well-defined short-range interaction, which when included in some appropriate quantum mechanical calculation will lead to the observed energy lowering between reactants and products. The mass-energy equivalence relation is only employed as a convenient means of computing this energy value numerically. There is never a suggestion in nuclear physics that $E = mc^2$ somehow describes the forces that are responsible for the observed binding. The same logic must apply to the electron-positron exothermic re action. The loss of mass is total in this case and this fact can be used conveniently to compute how much energy is expended in the process, but this relationship does not provide a satisfactory description of the forces that are at work in producing the enormous attraction between a particle and its antiparticle that is observed.

In the present study, arguments have been presented which lead to a concrete suggestion for such an interaction. Following Bohr's solution to the hydrogen atom problem, a potential is constructed which is characterized by a short-range minimum in total energy corresponding to the observed exothermicity of the electron-positron interaction, i.e. equal to $2m_ec^2$. It is pointed out that the Breit-Pauli Hamiltonian contains terms of order $\alpha^2 r^{-3}$ such as spin-orbit, spin-spin and orbit-orbit coupling which can account for the inner part of the corresponding potential well, but that some damping factor of even shorter range is needed to actually form the desired minimum and prevent total collapse of the $e^+$ $e^-$ system. The need for such an abrupt change in slope of the total energy curve near this minimum appears to be consistent with only an exponential type of damping, the details of which can be deduced from the nature of the Maxwell electromagnetic field Hamiltonian, specifically that the next term in the potential energy expansion should be of order $\alpha^4 r^{-4}$.

The resulting exponentially damped Hamiltonian is given explicitly in Table 1. Explicit calculations based on it have been carried out for the positron-electron and related systems and the results obtained will be discussed in a subsequent paper. The desired amount of binding is obtained when the constant A in the exponential damping factors has a value on the order of unity when atomic units are employed, as would be expected on the basis of the above qualitative arguments involving the classical electromagnetic Hamiltonian. Because the operator given in Table 1 is capable of treating both the long-range Coulomb effect as well as forces of much shorter range, it is possible to look upon it as a concretization of the electroweak interaction, and thus to anticipate that its range of applicability includes the description of neutron decay and other processes which are conventionally interpreted in terms of the creation and annihilation of particles of relatively small rest mass.

A scaling theorem can be derived based on the properties of the damped Breit-Pauli Hamiltonian in Table 1 which guarantees that the analogous potential lead to general particle-antiparticle binding energies which are proportional to the rest mass of each of the constituents. In this way the goal of having the nature of the interaction lead to results that are consistent with the Einstein mass-energy equivalence relation is achieved at least in principle. This result strengthens the view that particle-antiparticle interactions can be treated in the same manner as conventional binding processes in which the constituents are of different rest mass.

The nature of the interactions deduced in the above manner also helps to answer another important question that arises in the present interpretation, namely why there is no comparable state of exceptionally large binding energy for the proton-electron system. For any binary system in a state of zero translational energy, it follows that since $\mathbf{p}_1 + \mathbf{p}_2 = 0$, the respective velocities of the particles must be inversely proportional to their rest masses. This means that the speed of both particles must be equal in absolute magnitude for e+ e- and p+ p-, but they must differ by a factor of 1836 for the corresponding proton-electron system. Inspection of the exponentially damped Breit-Pauli Hamiltonian in Table 1 shows that a high degree of binding requires that the two interacting particles must spend a reasonably large fraction of time not only in close proximity to one another but also while each is moving at high speed. This pair of conditions is virtually never satisfied when the two particles must travel with widely different velocities as in the $p^+e^-$ case, whereas there is a good possibility that it can frequently be met in the particle-antiparticle case, since then the two velocities must remain equal (for p= 0) in absolute magnitude.

Although the present study has dealt almost exclusively with processes involving photons, it is clear that the main avenue of approach can be used to deal with a wide range of other topics in the field of particle physics as well. If the hypothesis of creation and annihilation of matter is false, it can only be replaced by the concept of complete elementary balance in all physical transformations, similarly as in the present interpretation of the decay of positronium.

To further illustrate the advisability of questioning the creation and annihilation hypothesis on a quite general basis, it is well to consider a problem of current interest that does not involve photons. For over thirty years now there has been experimental evidence [46] indicating that the solar neutrino flux arriving at the earth's surface is significantly smaller than what must be inferred from the standard solar model. It is important to recognize that the theoretical basis for this conclusion lies in the creation-annihilation hypothesis. Specifically, it is always assumed that for every antineutrino ($\bar{\nu}$) consumed in a particular fusion reaction a neutrino (ν) must be created with it, and it is on this basis that the expected neutrino flux is computed.

If one assumes instead that massless $\nu\bar{\nu}$ systems exist in infinite density throughout the universe, analogous to the E = 0 $e^+$ $e^-$ binaries discussed above, it becomes necessary to look upon the conventional production of a neutrino and antineutrino as involving dissociation of another tightly bound diatomic species. In this case, however, the minimum total energy of the particle-antiparticle system is the same as that of its separated constituents by virtue of the vanishingly small rest mass of the neutrino particles. Hence the $\nu\bar{\nu}$ binaries must exist in a metastable state that is subject to predissociation via a tunnelling process not available to either of its $e^+e^-$ or $p^+p^-$ counterparts. As a result it is reasonable to expect that a thermal equilibrium exists between such a system and its separated constituents, and thus that a certain fraction of the $\nu\bar{\nu}$ species can always be found in a dissociated state.

Whenever the solar fusion reactions occur with free antineutrinos, however, it is no longer necessary to assume that a companion neutrino is ejected with high velocity away from the scene. The fraction of dissociated $\nu\bar{\nu}$ species must be expected to be larger in the high-temperature solar environment than in typical nuclear reactors, so one may plausibly assume that their influence is far greater on the sun. To this can be added that the reactivity of free antineutrinos is much higher than those that need to be first produced in a conventional creation process. The Reines-Cowan experiment [47], for example, only becomes feasible when free (reactor) antineutrinos are employed. The fact that more recent experiments with a gallium detector [48] indicate an even higher neutrino deficiency than in previous experiments speaks for the fact that the production of neutrinos is very sensitive to the type of fusion process undergone, something which again is far more easily understood when one thinks of real $\nu\bar{\nu}$ binary systems being involved than in terms of the creation and annihilation of neutrinos from nothing. Such a solution to the solar neutrino puzzle is thus seen as a viable alternative to the proposal of neutrino oscillations [49].

Finally, one should avoid taking the position that since the theory of elementary particle physics is so far developed and relies so heavily on the creation and annihilation of matter hypothesis, that no alternative theory that denies the latter proposition could possibly be correct. Assuming that particles appear and disappear is extremely convenient from the standpoint of eliminating the need to arrive at a consistent description of the interactions which such systems must otherwise have undergone, as consideration of the proposal of a tightly bound $e^+$ $e^-$ complex has demonstrated quite clearly. It is therefore important to give careful consideration to the possibility that the fundamental building blocks of nature are as indestructible as the ancient Greeks assumed them to be, and that as a result, every physical transformation as yet observed can be described in a consistent manner by assuming that the number and types of elements involved is always the same on both sides of the corresponding reactive equation.

**Acknowledgment**

The author wishes to express his thanks to Dr. Robin A. Phillips and Mr. Peter Liebermann for their efforts in constructing computer programs and carrying out calculations which were instrumental in reaching the conc1usions of the present study. He also is especially grateful to Dr. Gerhard Hirsch for numerous helpful discussions during the course of this work and for his help in preparing the manuscript. The financial support of the Deutsche Forschungsgemeinschaft in the form of a Forschergruppe grant is also hereby gratefully acknowledged.

**References**

1. Dirac, P.A.M., 1931, *Proc. R. Soc. London A* 133, 80.
2. Anderson, C.D., 1933, *Phys. Rev.* 43, 491.
3. Chamberlain, O., Segre, E., Wiegand, C., and Ypsilantis T., 1955, *Phys. Rev.* 100, 947.
4. Einstein, A.,1905, *Ann. Physik* 17, 891; Lorentz, H.A., Einstein, A., Minkowski, H., and Weyl, H., 1952, *The Principle of Relativity* (Dover, New York), p. 35.
5. Bohr, N., 1913, *Phil. Mag.* 26, 1.
6. Buenker, R.J., 1992, *Molec. Phys.* 76, 277.
7. Fermi, E., and Yang, C.N., 1949, *Phys. Rev.* 76, 1739.
8. de Broglie, L., 1926, *Phil. Mag.* 47, 446; 1925, *Ann. Physik* 3, 22.
9. Davisson, C., and Germer, L.H., 1927, *Phys. Rev.* 30, 705.
10. Paul, H., 1985, *Photonen: Experimente und ihre Deutung* (Vieweg, Braunschweig), p. 27.
11. Jackson, J.D., 1962, *Classical Electrodynamics* (Wiley, New York).
12. Planck, M., 1901, *Ann. Physik* 4, 553.
13. Wigner, E.P., 1957, *Rev. Mod. Phys.* 29, 255.
14. Beth, R.A., 1936, *Phys. Rev.* 50, 115.
15. Allen, P.J., 1964, *Am. J. Phys.* 34, 1185.
16. Frauenfelder, H., and Henley, E.M., 1974, *Subatomic Physics* (Prentice-Hall, Englewood Cliffs, N.J.) , pp. 201-202.
17. Newton, 1., 1704, *Opticks* (London).
18. Weidner, R.T., and Sells, R.T., 1960, *Elementary Modern Physics* (Allyn and Bacon, Boston), pp. 35-37.
19. Einstein, A., 1951, Letter to M. Besso, cited in: Paul, H., 1985, *Photonen: Experimente und ihre Deutung* (Vieweg, Braunschweig), p. 7.
20. The German word Einstein actually used was "Lump", a humorously derogatory term more accurately translated as "rascal" or "scoundrel".
21. Compton, A.H., 1923, *Phys. Rev.* 21, 715; 22, 409.
22. Raman, C.V., and Krishnan, K.S., 1928, *Nature* 121, 501; 121, 619.
23. Smekal, A., 1923, *Naturwissenschaften* 11, 873.
24. Shortley, G. and Williams, D., 1961, *Elements of Physics,* Third edition (Prentice Hall, Englewood Cliffs, N.J.), pp. 871-872.
25. Goldstein, H., 1950, *Classical Mechanics* (Addison-Wesley, Reading, Mass.), p. 191.
26. Pauli, W., 1931, in: *Proceedings of the Meeting of the American Physical Society*

(Pasadena); 1933, in: *Proceedings of Solvay Congress* (Brussels), p. 324.

27. Morse, P.M., 1964, *Thermal Physics* (Benjamin, New York), pp. 350-353.
28. Schrödinger, E., 1926, *Ann. Physik* 79, 361; 79, 489; 80, 437; 81, 109.
29. Dirac, P.A.M., 1928, *Proc. R. Soc. London A* 117, 610.
30. Breit, G., 1929, *Phys. Rev.* 34, 553.
31. Pauli, W., 1927, *Z. Physik* 43, 601.
32. Bethe, H.A., and Salpeter, E.E., 1957, *Quantum Mechanics of One- and Two-Electron Atoms* (Springer, Berlin), p. 181.
33. Slater, J.C., 1960, *Quantum Theory of Atomic Structure,* Vol. 11 (McGraw-Hill, New York), p. 191.
34. Dirac, P.A.M., 1958, *The Principles of Quantum Mechanics,* Fourth edition (Clarendon Press, Oxford), p. 312.
35. Salam, A., 1968, in: *Elementary Particle Physics,* edited by N. Svartholm (Almquist and Wiksell, Stockholm), p. 367.
36. Georgi, H., and Glashow, S.L., 1972, *Phys. Rev. D,* 6, 2977.
37. Weinberg, S., 1974, *Rev. Mod. Phys.,* 46, 255.
38. Abers, E.S., and Lee, B.W., 1973, *Phys. Rept.* C, 9, 1.
39. Goeppert Mayer, M., 1948, *Phys. Rev.* 74, 235; 1949, 75, 1969; 1950, 78, 16.
40. Haxel, O., Jensen, J.H.D., and Suess, H., 1949, *Phys. Rev.* 75, 1766; 1950, *Z. Physik* 128, 295.
41. Feingold, A.M., 1952, *Thesis,* Princeton University.
42. Buenker, R.J., Chandra, P., and Hess, B.A., 1984, *Chemical Physics* 84, 1.
43. Bethe, H.A., and Salpeter, E.E., 1957, *Quantum Mechanics of One- and Two-Electron Atoms* (Springer. Berlin), p. 114.
44. Ore, A., and Powell, J., 1949, *Phys. Rev.* 75, 1696; 75, 1963.
45. Sachs, R.G., 1953, *Nuclear Theory* {Addison-Wesley, Cambridge, Mass.), p. 28.
46. Davis, Jr., R., Harmer, D.S., and Hoffmann, K.C., 1968, *Phys. Rev. Lett.* 20, 1205; Bahcall, J.N., 1964, *Phys. Rev. Lett.* 12, 300; Davis, Jr., R., 1964, *Phys. Rev. Lett.* 22, 303.
47. Reines, F., and Cowan, C.L., 1956, *Science* 124, 103; 1959, *Phys. Rev.* 113, 273; Reines, F., Cowan, C.L., Harrison, F.B., McGuire, A.D., and Kruse, H.W., 1960, *Phys. Rev.* 117, 159.
48. Bowles, T., 1990, *Soviet-American Gallium Experiment,* Seminar at Argonne National Laboratory.
49. Foukal, P., 1990, *Solar Astrophysics* (Wiley, New York), p. 196.

**TABLE I.** Definition of quantum mechanical operators present in the exponentially damped Breit-Pauli Hamiltonian discussed in the present study which has been identified with the electroweak. interaction ($\alpha$ is the fine-structure constant; atomic units employed throughout). The indices i and j are used generically to represent two interacting particles; the quantities $q_i$ ($q_j$) and $m_{0i}$ ($m_{0j}$) are the electric charges and rest masses of the ith (jth) particle, A is the exponential damping constant (units of $e^{-1}$; see text), and $p_i$ ($p_j$), $s_i$ ($s_j$) and $r_{ij}$ are the standard vectorial symbols for the linear and spin angular momenta of a single particle and the distance between the ith and jth particles respectively.

| Designation | | Operator |
|---|---|---|
| Relativistic Kinetic Energy (one-particle only) | KE | $((p_i^2 \alpha^{-2} + m_{0i}^2 \alpha^{-4})^{1/2} - m_{0i} \alpha^{-2}$ |
| Coulomb | C | $q_i q_j r_{ij}^{-1}$ |
| Spin-same-orbit (exponentially damped) | SsO | $-\dfrac{\alpha^2}{2} G(i,j)$ $\times \left\{ \left(\dfrac{q_i}{m_{0i}}\right)^2 \exp\left(-2A\alpha^2 \left\lvert\dfrac{q_i}{m_{0i}} p_i\right\rvert\right) \right.$ $\times (\mathbf{r_{ij}} \times \mathbf{p_i} \cdot \mathbf{s_i}) r_{ij}^{-3} \exp\left(-2A\alpha^2 \left\lvert\dfrac{q_i}{m_{0i}} p_i\right\rvert\right)$ $+ \left\{ \left(\dfrac{q_j}{m_{0j}}\right)^2 \exp\left(-2A\alpha^2 \left\lvert\dfrac{q_j}{m_{0j}} p_j\right\rvert\right) \right.$ $\times (\mathbf{r_{ji}} \times \mathbf{p_j} \cdot \mathbf{s_j}) r_{ij}^{-3} \exp\left(-2A\alpha^2 \left\lvert\dfrac{q_j}{m_{0j}} p_j\right\rvert\right),$ where $G(i,j) = \begin{cases} 1 \text{ if } \dfrac{q_i q_j}{m_{0i} m_{0j}} > 1 \\ -1, \text{ if } \dfrac{q_i q_j}{m_{0i} m_{0j}} < 1 \end{cases}$ |
| Spin-other-orbit (exponentially damped) | SoO | $-\alpha^2 \left(\dfrac{q_i}{m_{0i}}\right)\left(\dfrac{q_j}{m_{0j}}\right)$ $\times \exp\left(-A\alpha^2 \left\lvert\dfrac{q_i}{m_{0i}} p_i\right\rvert\right) \exp\left(-A\alpha^2 \left\lvert\dfrac{q_j}{m_{0j}} p_j\right\rvert\right)$ $\times (\mathbf{r_{ji}} \times \mathbf{p_j} \bullet \mathbf{s_i} + \mathbf{r_{ij}} \times \mathbf{p_i} \bullet \mathbf{s_j}) \, r_{ij}^{-3}$ $\times \exp\left(-A\alpha^2 \left\lvert\dfrac{q_i}{m_{0i}} p_i\right\rvert\right) \exp\left(-A\alpha^2 \left\lvert\dfrac{q_j}{m_{0j}} p_j\right\rvert\right)$ |

| Designation | | Operator |
|---|---|---|
| Darwin Term | D | $-\pi \dfrac{\alpha^2}{2} G(i,j)$ |
| (exponentially damped) | | $\times \left\{ \left(\dfrac{q_i}{m_{0i}}\right)^2 \exp\left(-2A\alpha^2 \left|\dfrac{q_i}{m_{0i}} p_i\right|\right) \right.$ |
| | | $\times \left\{ \delta(r_{ij}) \exp\left(-2A\alpha^2 \left|\dfrac{q_i}{m_{0i}} p_i\right|\right) \right.$ |
| | | $+ \left\{ \left(\dfrac{q_j}{m_{0j}}\right)^2 \exp\left(-2A\alpha^2 \left|\dfrac{q_j}{m_{0j}} p_j\right|\right) \right.$ |
| | | $\times \left\{ \delta(r_{ij}) \exp\left(-2A\alpha^2 \left|\dfrac{q_j}{m_{0j}} p_j\right|\right) \right.,$ |
| | | where $G(i,j) = \begin{cases} 1 \text{ if } \dfrac{q_i q_j}{m_{0i} m_{0j}} > 1 \\ -1, \text{ if } \dfrac{q_i q_j}{m_{0i} m_{0j}} < 1 \end{cases}$ |
| Orbit-orbit | OO | $-\dfrac{\alpha^2}{2}\left(\dfrac{q_i}{m_{0i}}\right)\left(\dfrac{q_j}{m_{0j}}\right)$ |
| (exponentially damped) | | $\times \exp\left(-A\alpha^2\left|\dfrac{q_i}{m_{0i}} p_i\right|\right) \exp\left(-A\alpha^2\left|\dfrac{q_j}{m_{0j}} p_j\right|\right)$ |
| | | $\times [(\mathbf{p_i} \bullet \mathbf{p_j}) r_{ij}^{-1} + (\mathbf{r_{ij}} \bullet (\mathbf{r_{ij}} \bullet \mathbf{p_i})\mathbf{p_j}\, r_{ij}^{-3}]$ |
| | | $\times \exp\left(-A\alpha^2\left|\dfrac{q_i}{m_{0i}} p_i\right|\right) \exp\left(-A\alpha^2\left|\dfrac{q_j}{m_{0j}} p_j\right|\right)$ |
| Spin-spin | SS | $\alpha^2\left(\dfrac{q_i}{m_{0i}}\right)\left(\dfrac{q_j}{m_{0j}}\right)$ |
| (exponentially damped) | | $\times \exp\left(-A\alpha^2\left|\dfrac{q_i}{m_{0i}} p_i\right|\right) \exp\left(-A\alpha^2\left|\dfrac{q_j}{m_{0j}} p_j\right|\right)$ |
| | | $\times [(\mathbf{s_i} \bullet \mathbf{s_j}) r_{ij}^{-3} - 3(\mathbf{r_{ij}} \bullet \mathbf{s_i})(\mathbf{r_{ij}} \bullet \mathbf{s_j}) r_{ij}^{-5}]$ |
| | | $\times \exp\left(-A\alpha^2\left|\dfrac{q_i}{m_{0i}} p_i\right|\right) \exp\left(-A\alpha^2\left|\dfrac{q_j}{m_{0j}} p_j\right|\right)$ |

TABLE I continued

| Designation | | Operator |
|---|---|---|
| Spin-spin δ | SSδ | $-\dfrac{8\pi\alpha^2}{3}$ |
| (exponentially damped) | | $\times \exp\left(-A\alpha^2 \left|\dfrac{q_i}{m_{0i}} p_i\right|\right) \exp\left(-A\alpha^2 \left|\dfrac{q_j}{m_{0j}} p_j\right|\right)$ |
| | | $\times \mathbf{s_i} \cdot \mathbf{s_j}\, \delta(r_{ij})$ |
| | | $\times \exp\left(-A\alpha^2 \left|\dfrac{q_i}{m_{0i}} p_i\right|\right) \exp\left(-A\alpha^2 \left|\dfrac{q_j}{m_{0j}} p_j\right|\right)$ |

**Figure Captions:**

Fig. 1: Energy level diagram comparing the hydrogen atom and positronium systems. In the standard quantum mechanical theory the lowest (1s) levels of each system occur at -0.5000 and -0.2500 hartree respectively. Corresponding ionization energies for the excited states of these two systems always differ by a factor of two as well, reflecting the different reduced masses of the electron in the two cases. The hydrogenic 1s state is known to be stable, however, whereas the corresponding $e^+e^-$ state has a short lifetime and decays radiatively. This suggests that the true lowest state of the $e^+e^-$ system actually lies far below the n=1 state of positronium, and as such corresponds to the massless state of the photon itself, with an "ionization" energy equal to $2m_{0e}c^2$ or 37557.7306 hartree.

Fig.2: Schematic diagram for the two-photon decay of (singlet) positronium. By assuming that the photon also has an $e^+e^-$ composition, it is possible to describe this transition without assuming that particles are either created or annihilated in the process. In this model three $e^+e^-$ binaries are involved, two of which are massless photons at the start of the process. They share the energy released by the positronium decay, and are observed as γ photons of equal energy at its conclusion. The final state of the original positronium system is another massless photon that, as its two counterparts at the start of the transition, escapes detection by virtue of its lack of energy.

Fig.3: Schematic diagram showing the variation of the internal energy of the $e^+e^-$ system as a function of the reciprocal of the distance between the two constituent particles. At large separations the Coulomb attractive interaction dominates because of its long-range ($r^{-1}$) character. A minimum energy of only -0.25 hartree is eventually reached, corresponding to the familiar hydrogenic 1s state of positronium, after which the total energy starts to rise because the shorter-range kinetic energy term begins to dominate. In the present model this trend is eventually reversed again at a much smaller $e^+e^-$ separation, at which point attractive forces of even shorter range ($\sim r^{-3}$) begin to change more rapidly than the kinetic energy. Finally, a second potential minimum, much deeper than the first, is reached which corresponds to a binding energy exactly equal to the sum of the rest energies of the electron and positron. At still smaller interparticle distances the total energy rises again, reflecting the effect of some momentum-dependent damping of the short-range

potential in this region. A form for the potential is sought for which only the Coulomb energy minimum survives when the positron is replaced by a proton.

Fig.4: Schematic diagram exploring the nature of a short-range potential required to produce the type of $e^+e^-$ total energy curve depicted in Fig. 3. At relatively small interparticle distances ($r \approx \alpha^2$) the relativistic kinetic energy varies nearly linearly with momentum $p \approx r^{-1}$. In order to obtain strong binding within a very narrow range of the $e^+$ - $e^-$ distance, the negative of the attractive potential term must reach a maximum shortly after it crosses the kinetic energy line from the long-distance side of the diagram, and then drop off again very sharply. Such an extreme cancellation effect requires a potential which fulfills at least three conditions: a) a small coupling constant (order $\alpha^2$), b) a shorter range ($\sim r^{-3}$) than the kinetic energy and c) a momentum-dependent damping which is exponential in nature.

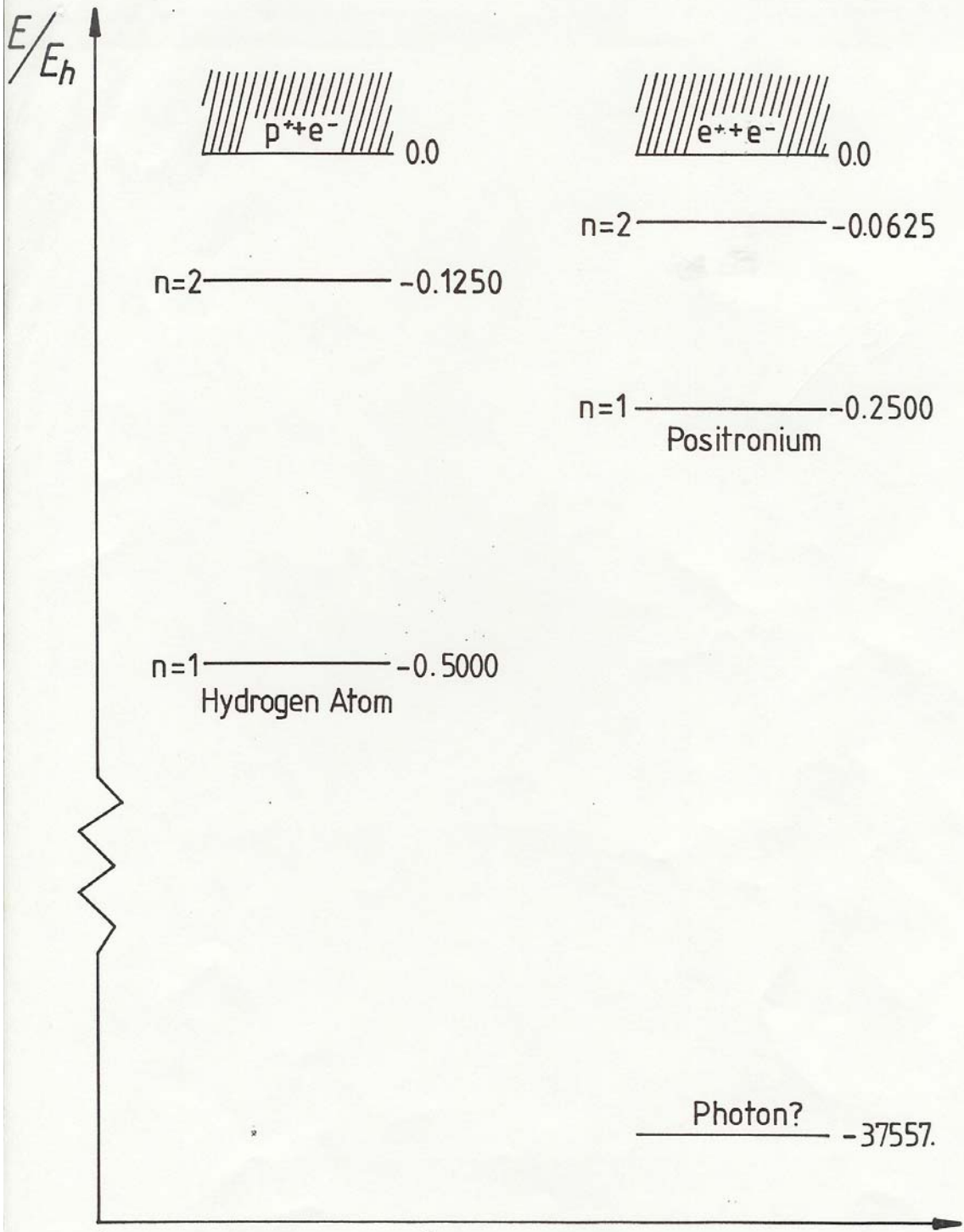

Figure 1
Buenker

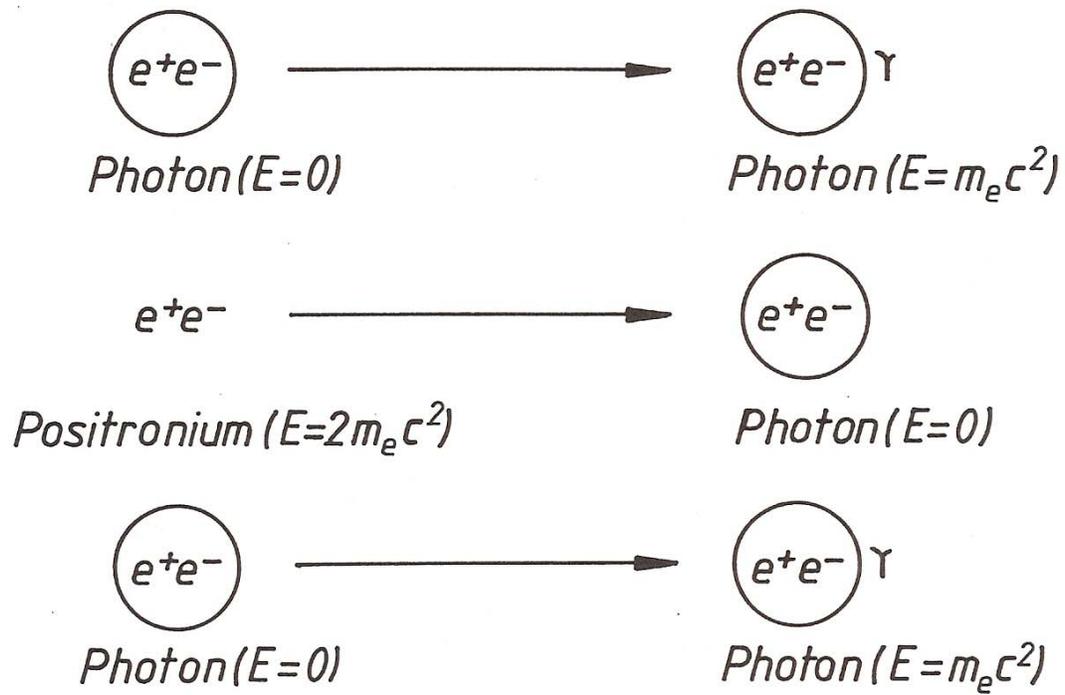

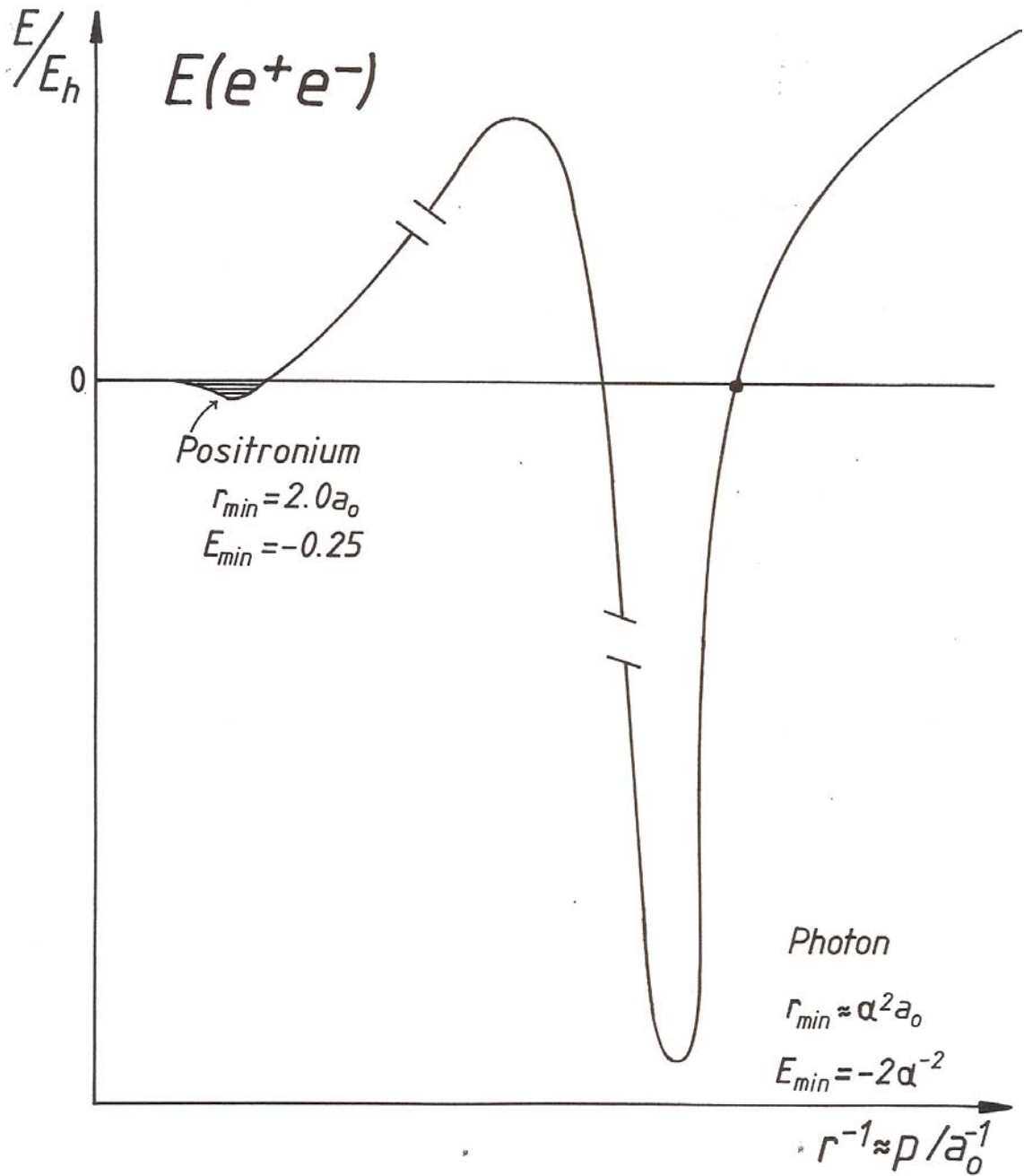

Figure 3
Buenker

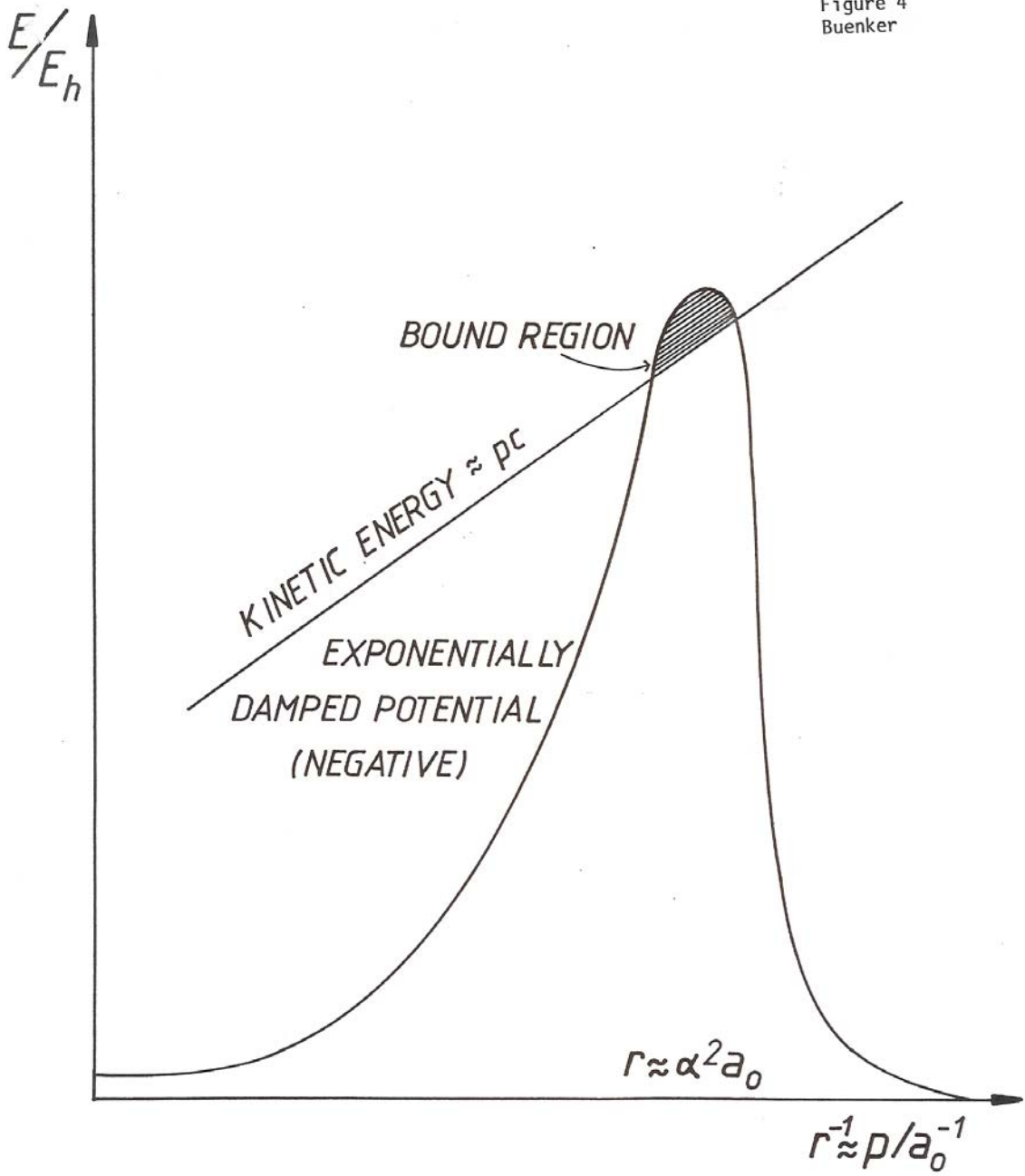

Figure 4
Buenker